# Don't Leave Me Alone: Retrospective Think Aloud supported by Real-time Monitoring of Participant's Physiology

Alexandros Liapis[1], Christos Katsanos[2,3] and Michalis Xenos[4]

[1] School of Science and Technology, Hellenic Open University, Patras, Greece
[2] Department of Informatics, Aristotle University of Thessaloniki, Thessaloniki, Greece
[3] HCI Group, Electrical and Computer Engineering Department, University of Patras, Patras, Greece
[4] Computer Engineering and Informatics Department, University of Patras, Patras, Greece
`aliapis@eap.gr,ckatsanos@ece.upatras.gr,xenos@upatras.gr`

**Abstract.** Think aloud protocols are widely applied in user experience studies. In this paper, the effect of two different applications of the Retrospective Think Aloud (RTA) protocol on the number of user-reported usability issues is examined. To this end, 30 users were asked to use the National Cadastre and Mapping Agency web application and complete a set of tasks, such as measuring the land area of a square in their hometown. The order of tasks was randomized per participant. Next, participants were involved in RTA sessions. Each participant was involved in two different RTA modes: (a) the strict guidance, in which the facilitator stayed in the background and prompted participants to keep thinking aloud based on his judgement and experience, and (b) the physiology-supported interventions, in which the facilitator intervened based on real-time monitoring of user's physiological signals. During each session, three participant's physiological signals were recorded: skin conductance, skin temperature and blood volume pulse. Participants were also asked to provide valence-arousal ratings for each self-reported usability issue. Analysis of the collected data showed that participants in the physiology-supported RTA mode reported significantly more usability issues. No significant effect of the RTA mode was found on the valence-arousal ratings for the reported usability issues. Participants' physiological signals during the RTA sessions did not also differ significantly between the two modes.

**Keywords:** Human-computer interaction, physiological signals, usability evaluation, retrospective think aloud.

## 1 Introduction

Software development industry has been increasingly focusing on usability as one of the most critical quality characteristics of an interactive system. Usability evaluation constitutes the key process to improve usability [1, 2]. Quantitative usability metrics, such as 'time on task' and 'task completion rate' provide a way to objectively evalu-ate the usability of an evaluated system [3], but fail to offer qualitative insight about the root of potential issues in the user experience [4]. On the other hand, qualitative approaches, such as questionnaires, interviews and video analysis, can provide such qualitative data, but these methods are prone to subjectivity and can be time consuming. More recently, researchers and practitioners have introduced new user experience evaluation approaches using facial expression, speech tone and keystroke analysis [5]. Collecting and analyzing data from users' physiology (e.g., heart rate, respiration, skin conductance) is also a powerful recent usability evaluation method [6–9].

Think-aloud (TA) protocol is a qualitative tool that is used to understand users' behavior while interacting with a system in the context of a usability evaluation study. TA protocol was originally developed to support researchers and practitioners in the domain of cognitive psychology for gaining insight into people's mental processes. Later, it was used to study users' performance in activities such as reading, writing and decision-making in various domains. The HCI field has also adopted the TA protocol, which is on the top of the usability evaluation list for many practitioners [10].

During a TA session, participants are required to verbalize their thoughts about their interaction experience, while they perform tasks on the evaluated system. This method enables evaluators to identify usability issues that need to be resolved in the next system version. Such usability issues may cause activation of users Autonomic Nervous System (ANS), which is known as the "fight or flight" response or stress [10, 11]. Computer users with frequently or daily exposure to stressors are in high risk to confront chronic stress, which may badly affect their health [12]. Apart from health issues, stress may also affect users' performance [13], and its presence in interactive computer environments is typically interpreted as a user experience issue.

According to Nielsen [1], TA is the most valuable single usability engineering method. It is a simple and useful technique for data collection, but it has been criticized [14] for noisy or inaccurate data, due to extra cognitive effort imposed on participants. In [15] two modes of TA application are proposed: "concurrent" and "retrospective". Both protocol modes are widely used by HCI researchers and practitioners. In the concurrent mode, participants are asked to verbalize their interaction experience, while working on the task. One main drawback of this mode is that it may affect the way that participants interact within the task, the time they need to complete the task, and their success in task completion [16]. The specific time cost is referred as reactivity effect.

In the retrospective mode, known as Retrospective Think Aloud (RTA), participants verbalize their interaction experience at the end of a task or a set of tasks. This is often done while viewing a recording of their interaction session. RTA, appears to yield more complex and explanatory data, as the test users who participate in the specific session are not under pressure; instead they are free to think aloud in a natural way [17]. Moreover, since the participants are free to perform the tasks without the need to think aloud, the risk of reactivity is eliminated. However, one of the most important drawbacks of the RTA method is that valuable segments of information may be lost due to participants' memory recall problem, as it has been confirmed by [18, 19]. Furthermore, RTA requires additional time, on top of the user testing session for both the participant and the facilitator.

The effectiveness of these two TA protocol modes in terms of usability issues detected has been examined [20, 21]. However, the effect of TA procedural aspects (i.e., when and how exactly a facilitator intervenes) on the effectiveness of the method remains rather unexplored. Ericsson and Simon [15] showed that application of TA strict guidance (i.e., a facilitator stays in the background just to prompt participants to keep thinking aloud) is very difficult to be applied. Therefore, they propose a free approach with more participant-facilitator interaction than the strict way.

The present study examines how two different treatments of the RTA protocol (a) strict guidance and (b) physiology-supported interventions, affect the number of the user-reported usability issues and users self-reported emotional ratings while experiencing these usability issues. In the strict guidance, condition the facilitator prompted participants to think aloud based on his judgement and experience. In the physiology-supported interventions condition, the same facilitator intervened based on real-time monitoring of user's physiological signals, such as skin conductance. In specific, the research questions investigated by this study are the following:

- RQ1: Is there any effect of the RTA mode on total number of usability issues reported by users in RTA sessions?
- RQ2: Is there any effect of the RTA mode on participants' self-reported ratings for their emotional state during a reported usability issue?
- RQ3: Is there any effect of RTA mode on participants' emotional state during the RTA session, as it is indicated by their physiological signals?

The rest of the paper is structured as follows. Section 2 presents the interaction tasks and the experimental general set-up and protocol, while in Section 3, the results from the experiment are presented. Finally, in Section 4, conclusions, limitations of the presented work and directions for future research are elaborated.

## 2  Interaction Scenarios and Experimental Setup

### 2.1  Scenarios

In this study, participants were asked to perform tasks using the free web-based Orthophotos viewing service[1] offered by the Greek National Cadastre and Mapping Agency (NCMA). In this web application, users can navigate the map of the whole country and perform tasks such as finding a specific place for a set of geographical coordinates, measuring distances on the map, measuring the area of a building etc. This web application was selected because a previous heuristic evaluation study, conducted by three experienced evaluators, showed that it has usability issues.

Participants were asked to use the service in order to perform two tasks. In the first task (see Fig.1 left), which included two sub-tasks, participants were asked to a) locate a well-known bridge in Patras (i.e.,. the bridge connecting Rio with Antirrio, known as 'Charilaos Trikoupis' bridge) and measure the distance between the first and the fourth pillar of this bridge and b) navigate in Patras old harbor and measure the length of the breakwater. In the second task (see Fig.1 right), which also involved two sub-tasks, participants were asked to a) locate a popular square (i.e., Georgiou Square) in the Patras city center and measure its inner area as defined by the dotted rectangle shown in the right part of Figure 1 and b) to modify the measured area to include all parts of the square as defined by the yellow polygon shown in the right part of Figure 1.

---

[1] http://gis.ktimanet.gr/wms/ktbasemap/default.aspx

None of the study participants had previous experience with the evaluated web application. However, most study participants (24/30) reported that they were rather experienced in map usage and navigation with such applications (e.g., Google Maps). Furthermore, the navigation places were carefully selected to be well-known to participants in an attempt to minimize the effect of spatial knowledge of the area on the interaction experience.

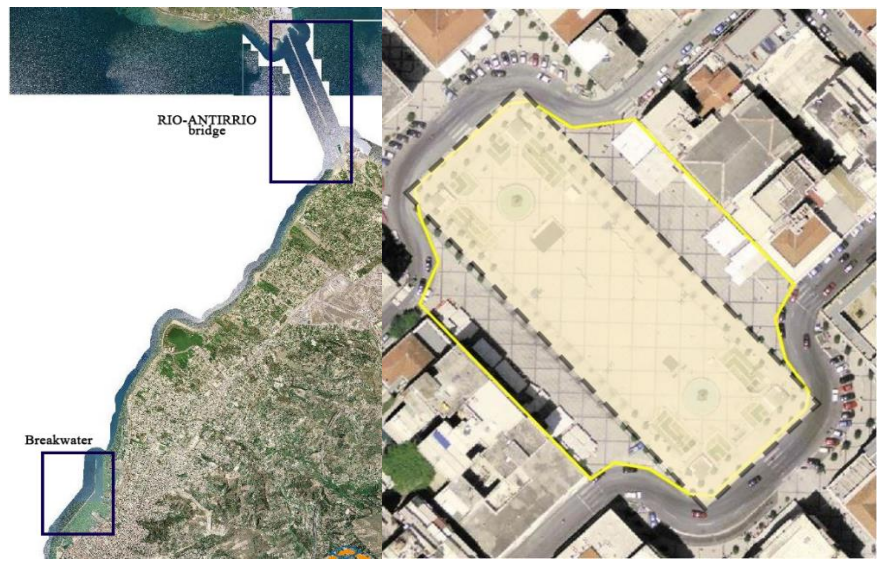

**Fig. 1.** The Orthophotos viewing service and specific spots used as part of the study tasks given to participants. Left: Participants were asked to find and measure distances (task 1). Right: Participants were asked to find and measure areas (task 2).

### 2.2 Experimental Setup

The experiment took place in the facilities of our fully-equipped usability lab. The wireless NeXus-10 physiological platform, along with BioTrace+ interface were used to manage physiological signals recording and real-time monitoring. Three physiological signals were recorded (skin conductance, skin temperature and blood volume pulse) with a sampling rate of 32Hz. All scenarios were designed to require minimum typing effort in order to minimize participants' hand movements that may affect physiological measurements.

During each experimental session, participants and facilitators were able to communicate through an intercom system. The desktop Tobii-studio recording environment was used to present the interaction scenarios to each participant. Level of room temperature and humidity were continuously monitored to minimize their effect on the collected physiological signals.

A sample of 30 healthy participants (17 males), aged between 18 and 45 (Mean=32.1, SD=7.1) was recruited. They were approached from university campus and the place of residence was the single criterion for their selection. Each experimental session lasted approximately 60 minutes, including short breaks between scenarios. At the end of the experiment, each participant was debriefed about study's purpose and access to their data sources (e.g., eye-activity and physiological signals) was offered as an option to them.

At the beginning of each experimental session, participants were informed that they will be asked to interact with an online map-based service in order to perform two tasks. Next, they completed an appropriate consent form, along with some demographic information. Afterwards, the physiological sensors were placed on participants' non-dominant hand. A short time of approximately five minutes was given to them in order to get used to the sensors' presence, while sensors' transmission quality and participant's body posture in front of the eye-tracker were checked by the experiment facilitators.

Before each task a two minutes relaxing video was presented to participants while their baseline of their physiological signals was recorded. Subsequently, scenarios were presented to participants in a counterbalance mode, in order to remove potential confounds during data analysis phase. At the end of the user testing session, participants answered the Greek version of the standardized 50-item Big Five Trait Test questionnaire[2]. The Google Forms service was used to implement the questionnaire and to collect participants' responses. However, the analysis of both the eye-tracking data and the Big Five ratings are beyond the scope of this paper.

After the user testing sessions, participants were engaged in a RTA session. RTA was applied in two different modes and it was supported by the PhysiOBS tool. PhysiOBS (see Fig.2) is an innovative tool that effectively combines observation data and self-reported data for continuous and emotional states analysis and is delineated in [22]. In this study, it was used to present the video of the user testing session to participants, create Areas of Interest (AOIs) indicating usability issues based on participant's thinking aloud, and assign participant's self-reported ratings to these AOIs.

During each RTA session, participants watched their corresponding interaction session (screen recording) through PhysiOBS. In the strict guidance RTA mode, the facilitator asked from participants to think aloud about their interaction experience and had no further involvement in the process, except reminding them to think aloud in cases of long pauses. In the physiology-supported interventions RTA mode, the facilitator was more engaged in the process. In specific, real-time monitoring of user's physiology (e.g., rising of skin conductance) served as an intervention mechanism for facilitator's actions, such as encouraging participants to think aloud or engaging in brief discussions related to the user-reported issue. The type of RTA session (i.e., strict guidance vs. physiology-supported interventions) was randomly assigned for each participant-task combination.

---

[2] http://ipip.ori.org/Greek50-itemBigFiveFactorMarkers.htm

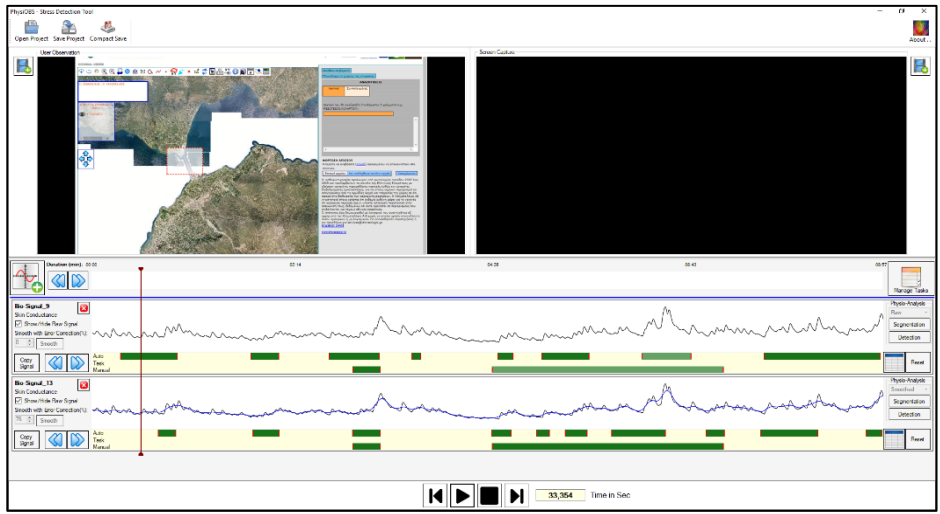

**Fig. 2.** PhysiOBS: A tool that supports synchronous viewing of multiple user experience data and users' emotional experience evaluation.

In addition, the Affect Grid [23] tool was used by participants in order to rate their emotional state in every usability issue they reported. The Affect Grid requires participants to select a point on a 9×9 grid that best indicates their emotional state associated with a stimulus, such as a usability issue. Grid's horizontal axis represents the valence (displeasure-pleasure) and the vertical axis the arousal (sleepiness-arousal). For example, if someone feels neutral, then the middle square of the grid (coordinates = 5, 5) is expected to be selected.

## 3   Analysis and Results

Thirty users participated in a study investigating the effect of two different applications of the RTA protocol on the number of reported usability issues and users self-reported emotional ratings while experiencing these usability issues. All in all, data from twenty-four participants (14 males), aged between 18 and 45 (Mean=32.3, SD=7.5) were analyzed. Six cases were excluded from analysis due to missing data (e.g., physiological data recording error). This was a within-subjects study and thus the data analysis was performed on 48 interaction sessions (24 participants x 2 RTA modes). In all subsequent statistical analyses, the effect size r was calculated according to the formulas reported in [24].

### 3.1 RQ1: RTA Mode and Total Number of User-Reported Usability Issues

Participants reported a total of 115 usability issues: 51 in strict guidance RTA mode and 64 in physiology-supported interventions RTA mode. No grouping was applied to produce a unique list of usability issues.

A two-tailed dependent samples t-test showed a significant difference in the number of usability issues that had been reported between strict guidance (M=2.08, SD=1.10) and physiology-supported interventions (M=2.96, SD=1.49) mode; t(23)=2.26, p=0.033, r=0.43. This medium-to-large observed effect size [25] demonstrates the importance of the RTA application mode on the effectiveness of the method in identifying usability issues. In specific, participants in the physiology-supported interventions RTA mode reported significantly more usability issues. A parametric test was used, because a Shapiro-Wilk test revealed that the distribution of the differences in the number of usability issues found by the two RTA modes did not deviate significantly from a normal distribution; W(24) = 0.95, p = 0.23.

### 3.2 RQ2: RTA Mode and VA Ratings for User-Reported Usability Issues

During usability issues reporting, participants were also asked to provide a rating of their emotional state using the valence-arousal space for each usability issue that they reported.

Figure 3 illustrates the valence and arousal ratings (N=115) for each usability issue in each RTA mode. The size of the bubble represents the number of ratings for each valence-arousal pair. Shapiro-Wilk test revealed that emotional ratings were not normally distributed (p<0.05) for both levels of the valence and arousal dependents. A non-parametric test (Mann-Whitney) showed that valence and arousal ratings were not significantly different between RTA modes; valence: Z=0.81, p=0.420, and arousal: Z=0.98, p=0.325.

The dotted frame in Figure 3 represents usability issues that caused intense emotions (Valence<5 and Arousal>5), such as stress [26]. Participants assigned more usability issues (N=33) in this area during the physiology-supported interventions than during the strict guidance (N=22) RTA mode. Shapiro-Wilk test revealed that emotional ratings within the stress area were not normally distributed (p<0.01) for both levels of the valence and arousal dependents. A Mann-Whitney test found no effect of RTA mode on valence and arousal ratings in the stress area; valence: Z=1.01, p=0.311, and arousal: Z=1.31, p=0.190.

### 3.3 RQ3: RTA Mode and Participants' Emotional State during the Thinking Aloud Session

Mean values of participants' physiological signals were recorded during each RTA session and were used as indicators of their emotional state. Table 1 presents descriptive statistics of participants' physiological signals during the two RTA modes.

Two-tailed Wilcoxon signed ranks tests found no significant difference (p>0.05) between the two RTA modes for the recorded signals; skin conductance: Z=0.64,

p=0.523, skin temperature: Z=0.63, p=0.530, and blood volume pulse: Z=0.16, p=0.875. Non-parametric tests were used because the assumption of normality was violated for all three recorded signals; skin conductance: W(24)=0.35, p<0.001, skin temperature: W(24)=0.78, p<0.001, and blood volume pulse: W(24)=0.88, p=0.010.

**Table 1.** Descriptive statistics of the physiological signals during the two RTA modes. GSR: Galvanic Skin Response, TEMP: Skin Temperature, BVP: Blood Volume Pulse.

| Signal | RTA mode | Mean | Median | SD | 95% CI |
|---|---|---|---|---|---|
| GSR | Strict | 4.05 | 2.35 | 4.31 | [2.23, 5.87] |
| GSR | Physiology-Supported | 3.75 | 2.53 | 3.16 | [2.42, 5.09] |
| TEMP | Strict | 28.43 | 29.58 | 5.68 | [27.28, 30.83] |
| TEMP | Physiology-Supported | 27.96 | 28.53 | 5.68 | [26.8, 30.36] |
| BVP | Strict | -22.91 | -20.06 | 12.34 | [-28.12, -17.70] |
| BVP | Physiology-Supported | -23.18 | -20.91 | 12.53 | [-28.5, -17.9] |

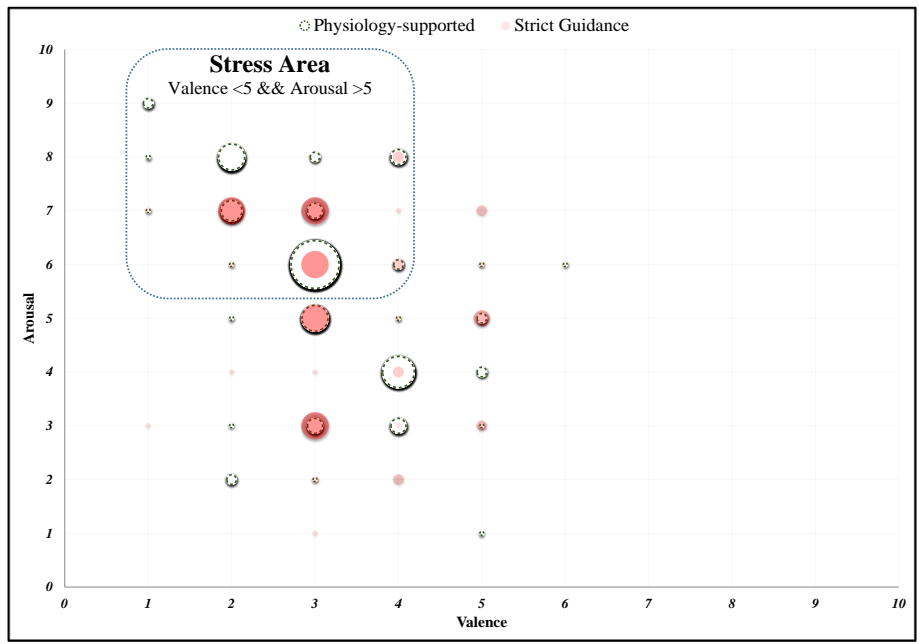

**Fig. 3.** Valence-Arousal ratings per usability issue for each RTA mode. Bubble size represents the number of ratings for each valence-arousal pair.

## 4     Conclusions, Limitations and Future Goals

The aim of this study is to provide usability researchers and practitioners with a better understanding of users' treatment during a RTA session. Think aloud is a popular testing method in collecting usability data. Studies like this one can help evaluators to make more informed decisions about RTA protocol application.

The results of this study demonstrate that the physiology-supported interventions RTA mode significantly outperformed the strict guidance RTA mode in terms of the number of usability issues reported by users. Participants' valence-arousal ratings for the reported usability issues did not differ significantly between the two RTA modes examined in this study. In addition, there was no effect of RTA mode on participants' physiological signals during the RTA sessions. However, in the physiology-supported interventions RTA mode, participants tended to report more stressful usability issues and to have lower mean values for all recorded physiological signals during the RTA sessions.

In sum, the physiology-supported interventions RTA mode seems to be the more appropriate method for evaluators who are interested in detecting more usability issues, rather than the typical strict guidance RTA mode.

As with any research, this study is not without limitations. First, the present study used a within-group design. Hence, no individual differences, such as personality traits and gender, and their possible effects on think-aloud performance was examined. To this end, we are already planning future similar experiments to extend the data collected in this study. In addition, there was one test moderator, and this person was the same between the two conditions (strict guidance and physiology-supported interventions). Although, this approach was chosen to ensure consistency across all users in each condition, it might have affected the results. Future studies need to engage more moderators and investigate this effect, if any. Furthermore, regarding the physiology-supported interventions condition, the interventions were based on real time visual inspection of the physiological signals. The use of an automatic mechanism (a kind of silent alert available only to facilitator) could probably be a parameter for further investigation in an attempt to make more systematic the triggering events of these interventions. Finally, future work also involves designing PhysiOBS-mediated learning activities for instruction of thinking aloud protocols and physiological monitoring of study participants in the context of our previous work [27–31] in HCI education.